\documentclass{iau}

\usepackage{amsmath}
\usepackage{graphicx}
\usepackage{multirow}
\usepackage{natbib}
\usepackage{bm}
\usepackage{mathrsfs}
\usepackage{xcolor}

\graphicspath{{./fig/}{./png/}}

\def\araa{ARA\&A}               
\def\apj{ApJ}                   
\def\apjl{ApJ}                  
\def\apjs{ApJS}                 
\def\aap{A\&A}                  

\def\mnras{MNRAS}               

\def\ssr{Space~Sci.~Rev.}       
\def\zap{ZAp}                   

\def\jgr{J.~Geophys.~Res.}      

\def\memsai{Mem.~Soc.~Astron.~Italiana}
\newcommand{\eee}{{\bm e}}

\newcommand{\uuu}{{\bm u}}

\newcommand{\FFF}{{\bm F}}

\newcommand{\gggg}{{\bm g}}
\newcommand{\ueeez}{\hat{\bm e}_z}
\newcommand{\Eq}[1]{Eq.~(\ref{#1})}

\newcommand{\EQ}{\begin{equation}}
\newcommand{\EN}{\end{equation}}
\newcommand{\EQA}{\begin{eqnarray}}
\newcommand{\ENA}{\end{eqnarray}}

\newcommand{\pd}{\partial}

\newcommand{\mean}[1]{\overline{#1}}

\newcommand{\cP}{c_{\rm P}}
\newcommand{\cV}{c_{\rm V}}
\newcommand{\cs}{c_{\rm s}}

\newcommand{\urms}{u_{\rm rms}}
\newcommand{\uconv}{u_{\rm conv}}
\newcommand{\ustar}{u_\star}

\newcommand{\Ma}{{\rm Ma}}

\newcommand{\kmax}{k_{\rm max}}
\newcommand{\tkmax}{\tilde{k}_{\rm max}}
\newcommand{\lmax}{\ell_{\rm max}}

\newcommand{\kmean}{k_{\rm mean}}
\newcommand{\tkmean}{\tilde{k}_{\rm mean}}

\newcommand{\lmean}{\ell_{\rm mean}}

\newcommand{\kH}{k_{\rm H}}

\newcommand{\Lx}{L_{\rm x}}
\newcommand{\Ly}{L_{\rm y}}
\newcommand{\Lz}{L_{\rm z}}

\newcommand{\lconv}{\ell_{\rm conv}}
\newcommand{\chiSGS}{\chi_{\rm SGS}}

\newcommand{\Co}{{\rm Co}}

\newcommand{\Col}{{\rm Co}_\ell}
\newcommand{\CoF}{{\rm Co}_{\rm F}}

\newcommand{\Hp}{H_{\rm p}}

\newcommand{\Pe}{{\rm Pe}}

\newcommand{\PraSGS}{{\rm Pr}_{\rm SGS}}

\newcommand{\RaFS}{{\rm Ra}_{\rm F}^\star}

\newcommand{\Rey}{{\rm Re}}

\newcommand{\Ta}{{\rm Ta}}

{}

\newcommand{\Omsun}{\Omega_\odot}

\newcommand{\Rsun}{R_\odot}

\newcommand{\Ftot}{F_{\rm tot}}

\newcommand{\Fn}{\mathscr{F}_{\rm n}}
\newcommand{\mFenth}{\mean{F}_{\rm enth}}
\newcommand{\mFconv}{\mean{F}_{\rm conv}}
\newcommand{\mFrad}{\mean{F}_{\rm rad}}
\newcommand{\mFkin}{\mean{F}_{\rm kin}}

\newcommand{\mFcool}{\mean{F}_{\rm cool}}

\newcommand{\mFD}{\mean{F}_{\rm D}}
\newcommand{\mFG}{\mean{F}_{\rm G}}
\newcommand{\Lsun}{L_\odot}
\newcommand{\nabad}{\nabla_{\rm ad}}
\newcommand{\nabD}{\nabla_{\rm D}}

\newcommand{\rbot}{r_{\rm bot}}

\newcommand{\ddz}{d_{\rm DZ}}
\newcommand{\tddz}{\tilde{d}_{\rm DZ}}

\newcommand{\dmix}{d_{\rm mix}}
\newcommand{\tdmix}{\tilde{d}_{\rm mix}}
\def\onethird{{\textstyle{1\over3}}}

\def\onehalf{{\textstyle{1\over2}}}

\newcommand{\Fig}[1]{Figure~\ref{#1}} 
\newcommand{\Figu}[1]{Figure~\ref{#1}}

\definecolor{ForestGreen}{RGB}{34,139,34}
\definecolor{AGray}{rgb}{.4,.4,.4}
\definecolor{LightYellow}{rgb}{1.,1.,.8}
\definecolor{LightCyan}{rgb}{0.88,1,1}

\begin{document}

\lefttitle{Petri J. K\"apyl\"a}
\righttitle{Effects of rotation and surface forcing on deep stellar convection zones}

\jnlPage{1}{7}
\jnlDoiYr{2021}
\doival{10.1017/xxxxx}

\aopheadtitle{Proceedings IAU Symposium No. 365}
\editors{Alexander Getling \&  Leonid Kitchatinov, eds.}

\title{Effects of rotation and surface forcing on deep stellar convection zones}

\author{Petri J.\ K\"apyl\"a}
\affiliation{Leibniz-Insitute for Solar Physics (KIS), Sch\"oneckstra{\ss}e 6, 79104 Freiburg im Breisgau, Germany}

\begin{abstract}
The canonical undestanding of stellar convection has recently been put
under doubt due to helioseismic results and global 3D convection
simulations. This ``convective conundrum'' is manifested by much
higher velocity amplitudes in simulations at large scales in
comparison to helioseismic results, and the difficulty in reproducing
the solar differential rotation and dynamo with global 3D
simulations. Here some aspects of this conundrum are discussed from
the viewpoint of hydrodynamic Cartesian 3D simulations targeted at
testing the rotational influence and surface forcing on deep
convection. More specifically, the dominant scale of convection and
the depths of the convection zone and the weakly subadiabatic -- yet
convecting -- Deardorff zone are discussed in detail.
\end{abstract}


\maketitle

\section{Introduction}

The solar convective envelope rotates differentially, such that the
rotation rate at the equator is about 40 per cent faster than at near
the poles. Furthermore, helioseismology has revealed that the angular
velocity $\Omega$ increases (decreases) with radius near the equator
(high latitudes), with narrow shear layers at the base and near the
surface of the convection zone \citep[e.g.][]{TCDMT03}. This
large-scale phenomenon is one of the principal observations that
global 3D simulations seek to reproduce. Early 3D simulations of the
late 1970s and early 1980s were able to capture this
\citep[e.g.][]{Gi77}, although dynamo cycles in those simulations did
not match that of the Sun \citep[e.g.][]{Gi83,Gl85}. However, it took
another two decades for such simulations to become more mainstream
\citep[e.g.][]{BMT04,GCS10,BMBBT11,KMB12}; see also
\cite{2023SSRv..219...58K} for a recent review. Soon thereafter it was
realized that obtaining solar-like differential rotation (fast
equator, slow poles) with simulations with the nominal solar rotation
rate and luminosity is highly non-trivial
\citep[e.g.][]{GYMRW14,KKB14,FF14,2016AdSpR..58.1475O}. This is
thought to be due to too weak rotational influence on the dominant
convective scales, or equivalently, a too low Coriolis (inverse
Rossby) number.

At the same time, efforts were made to study the velocity amplitudes
in the Sun using helioseismology
\citep{2010ApJ...712L..98H,HDS12}. These studies led to the
realization that convective amplitudes at horizontal scales of the
order of hundreds of Mm in the Sun appear to be several orders of
magnitude weaker than in the global simulations, and that the velocity
power spectrum in the Sun peaks at supergranular scale of
$20$-$30$~Mm. While the difference between helioseismic and simulation
results has reduced somewhat in the meantime, a large discrepancy
remains \citep[e.g.][]{2021PhDT........26P}. Adding to the puzzle are
the results of \cite{GHFT15} from a ring-diagram analysis that shows
high velocity amplitudes in the near-surface shear layer of the Sun
consistent with global 3D convection simulations.

Several physical processes have been suggested as possible solutions
of the convective conundrum. Rotationally constrained convection in
the deep parts of the convection zone is one such possibility
\citep{FH16b,2021PNAS..11822518V}. Linear stability analysis and
non-linear simulations of convection indicate that the convective
scale decreases with rotation. Given that the velocity power spectrum
peaks at the supergranular scale in the Sun, it is has been
conjectured that this scale coincides with the largest convectively
driven scale in the deep convection zone. The question if convection
in the Sun is indeed sufficiently constrained by rotation was studied
systematically in \cite{2023arXiv231012855K}. These results are
reviewed in more detail below.

Another possibility is that the solar convection zone is in fact
largely subadiabatic, that is, the thermal stratification is formally
weakly Schwarzschild stable. This can be enabled by plumes originating
near the surface that transport cool low entropy material deep into
the interior far beyond the formally unstable layer. This is related
to the idea that convection in the Sun is driven by the cooling at the
surface rather than by a superadiabatic temperature gradient
throughout the convection zone \citep{SN89,Sp97}. Such non-local
driving of convection due to surface cooling has been dubbed ``entropy
rain'' \citep[e.g.][]{Br16}. The convective flux in the stably
stratified, but convecting, layer is carried by a counter-gradient
term proportional to the variance of entropy fluctuations
\citep{1961JAtS...18..540D,De66}. Hence this layer is referred to as
the Deardorff zone. Simulations of overshooting convection routinely
capture such subadiabatic layers if the transition between the
radiative and convective regions is smooth enough
\cite[e.g.][]{1993A&A...277...93R,2015ApJ...799..142T,2017ApJ...845L..23K,2017ApJ...843...52H}. Most
of the previous works considered non-rotating cases, whereas here
recent results of \cite{2023arXiv231012855K}, where the effects of
rotation were included, are discussed.

Finally, the strength of the surface forcing depends on the physics
near the surface of the star. In real stellar convection zones the
density drops vertigineously near the surface and this cannot be
directly reproduced in numerical simulations
\citep[e.g.][]{2017LRCA....3....1K,2023SSRv..219...58K}. Here
preliminary results from an effort to study the effects of surface
forcing by varying the (imposed) surperadiabatic temperature gradient
at the surface are discussed based on earlier models presented in
\cite{2017ApJ...845L..23K}. The novelty of these simulations is that
they are constructed in such a way that the depth and structure of the
convection zone are self-consistent results of the models instead of
being fixed from the outset.

\section{The model}

The set-up is the same as in \cite{2019A&A...631A.122K},
\cite{2021A&A...655A..78K}, and \cite{2023arXiv231012855K}, and the
     {\sc Pencil Code} \citep{2021JOSS....6.2807P} was used to make
     the simulations. The simulation domain is a rectangular box with
     dimensions ($\Lx, \Ly, \Lz) = (4,4,1.5)d$, where $d$ is the depth
     of the initially isentropic layer which is situated between
     $0\leq z/d\leq1$. Initially this layer is sandwiched between a
     radiative layer with polytropic index $n=3.25$ ($-0.45\leq z/d
     <0$) and an isothermal layer ($1<z/d \leq1.05$). The equations
     for compressible hydrodynamics are solved:
\begin{eqnarray}
\frac{D \ln \rho}{D t} &=& -\bm\nabla \bm\cdot \uuu, \label{equ:dens}\\
\frac{D\uuu}{D t} &=& \gggg -\frac{1}{\rho}(\bm\nabla p - \bm\nabla \bm\cdot 2 \nu \rho \bm{\mathsf{S}}) - 2\bm\Omega \times \uuu,\label{equ:mom} \\
T \frac{D s}{D t} &=& -\frac{1}{\rho} \left[\bm\nabla \bm\cdot \left(\FFF_{\rm rad} + \FFF_{\rm SGS}\right) - \mathcal{C} \right] + 2 \nu \bm{\mathsf{S}}^2,
\label{equ:ent}
\end{eqnarray}
where $D/Dt = \pd/\pd t + \uuu\bm\cdot\bm\nabla$ is the advective
derivative, $\rho$ is the density, $\uuu$ is the velocity, $\gggg =
-g\ueeez$ with $g>0$ is the acceleration due to gravity where $\ueeez$
is the unit vector along the vertical ($z$) direction, $p$ is the gas
pressure, $\nu$ is the viscosity, $\bm{\mathsf{S}}$ is the traceless
rate-of-strain tensor, $\bm\Omega = \Omega\ueeez$ is the rotation
vector, $T$ is the temperature, and $s$ is the specific
entropy. $\FFF_{\rm rad} = - K \bm\nabla T$ is the radiative flux
where $K = K_0 \rho^{-2} T^{6.5}$ is the heat conductivity following
Kramers opacity law, and $\FFF_{\rm SGS} = - \chiSGS \rho T \bm\nabla
s'$ is the subgrid-scale (SGS) entropy flux, where $\chiSGS$ is a
constant SGS diffusivity and $s' = s -\mean{s}$ is the deviation of
the entropy from its horizontally averaged profile which is denoted by
the overbar. The gas obeys the ideal gas equation $p=\mathcal{R}\rho
T$, where $\mathcal{R}=\cP-\cV$ is the gas constant and where $\cP$
and $\cV$ are the specific heats in constant pressure and volume,
respectively. Finally, $\mathcal{C}$ describes cooling near the
surface.

In \cite{2023arXiv231012855K} it was shown that a Coriolis number
based on a hypothetical velocity $\ustar = (\Ftot/\rho)^{1/3}$ is
equivalent to
\begin{eqnarray}
\CoF = 2\Omega \Hp \left(\frac{\rho}{\Ftot} \right)^{1/3} = (\RaFS)^{-1/3},\label{equ:CoF}
\end{eqnarray}
where $\Ftot$ is the total energy flux, $\Hp = ({\rm d}\ln
\mean{p}/{\rm d} z)^{-1}$ is the pressure scale height, and
\begin{eqnarray}
\RaFS = \frac{g \Ftot}{8\cP \rho T \Omega^3 H^2} = \frac{\Ftot}{8\rho \Omega^3 \Hp^3},
\end{eqnarray}
is the flux-based diffusion-free modified Rayleigh number
\citep[e.g.][]{2002JFM...470..115C}, where the length scale $H$ was
chosen such that $H = \cP T/g = \Hp$, where $\Hp$ is taken at the base
of the convection zone. The advantage of $\CoF$ is that it does not
depend on any dynamical velocity or length scale and it can be
computed using observables $(\Omega, \Ftot)$ and quantities from
stellar structure models $(\rho, \Hp)$; see also the discussion in
\cite{2023A&A...669A..98K}. Further system parameters include the
Taylor number $\Ta = 4\Omega^2 d^4/\nu^2$, and the Prandtl number
related to the SGS diffusivity $\PraSGS=\nu/\chiSGS$. The energy flux
is measured by the dimensionless flux $\Fn=\Ftot/\rho\cs^3$ at
$z/d=-0.45$ in the initial non-convecting state. Diagnostic quantities
include the Reynolds ($\Rey=\urms/\nu k_1$) and P\'eclet number
($\Pe=\urms/\chiSGS k_1 = \PraSGS \Rey$), and the global Coriolis
number $\Co=2\Omega/(\urms k_1)$, where $\urms$ is the volume-averaged
rms-velocity, and $k_1=2\pi/d$ is an estimate of the scale of the
largest eddies. A more detailed description of the model is given in
\cite{2023arXiv231012855K}.

Three main sets of runs (Sets~A, B, and C) were made where $\Co$ was
varied between $0$ and about $17$. The imposed flux $\Fn$ was varied
between the sets to study the scaling of dynamical quantities with
respect to it. The diffusivities were varied proportional to
$\Fn^{1/3}$ to achieve the same $\Rey$, $\Pe$, and $\Co$ in each set
\citep[cf.][for more details]{2020GApFD.114....8K}. The primary
difference between the sets is that the Mach number $\Ma=\urms/\cs$,
where $\cs$ is the sound speed, and therefore relative stability of
the radiative layer below the convection zone vary. In Sets~A to C,
$\Rey=\Pe\approx 30\ldots 40$ and $\PraSGS=1$. A subset of Set~A,
denoted as Set~Am, was repeated at a higher resolution ($576^3$
instead of $288^3$ grid points), and correspondingly higher values of
Reynolds and P\'eclet numbers ($\Rey=\Pe\approx 65\ldots 84$), while
keeping $\RaFS$ fixed.

\section{Rotational scaling of convection}

\begin{figure}[t]
  \centering
  \includegraphics[scale=.5]{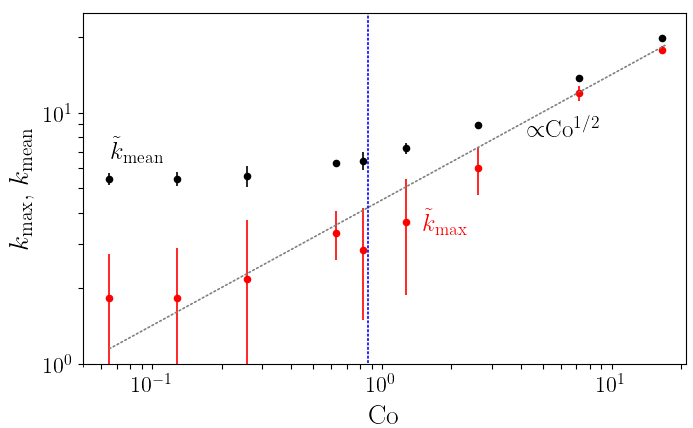}
  \caption{Mean ($\kmean$) and peak ($\kmax$) wavenumbers from power
    spectra of velocity as a function of $\Co$ from simulations Set~A
    from \cite{2023arXiv231012855K}. Data taken near the surface of
    the convectively unstable layer at $z/d = 0.85$. The dotted line
    shows the $\Co^{1/2}$ prediction from the CIA balance. The blue
    vertical dotted line indicates the solar value of $\CoF$. The
    tildes indicate normalization by $\kH$.}
  \label{fig:kmaxmmean}
\end{figure}

In \cite{2023arXiv231012855K} the scaling of various quantities in
rotating convection were studied. The numerical results were compared
with scalings derived for slow rotation where a balance between
inertial and buoyancy forces is assumed and for rapid rotation where a
balance between Coriolis, inertial, and Archimedean (buoyancy) forces,
or the CIA balance
\citep[e.g.][]{1979GApFD..12..139S,2014ApJ...791...13B,2020PhRvR...2d3115A},
is assumed. For the dominant convective scale this leads to:
\begin{eqnarray}
\lconv \sim \Hp\ \ \mbox{(slow rotation)},\ \ \mbox{and}\ \ \lconv \sim \Hp \Co^{-1/2}\ \ \mbox{(rapid rotation)}.\label{equ:lscaling}
\end{eqnarray}
Similarly, the scalings for the convective velocity are:
\begin{eqnarray}
\uconv \sim \ustar\ \ \mbox{(slow rotation)},\ \ \mbox{and}\ \ \uconv \sim \ustar \Co^{-1/6}\ \ \mbox{(rapid rotation)}.\label{equ:uscaling}
\end{eqnarray}
Finally, the local Coriolis number $\Col=2\Omega\lconv/\uconv$, can be
shown to depend on $\RaFS$:
\begin{eqnarray}
\Col \sim (\RaFS)^{-1/3}\ \ \mbox{(slow rotation)},\ \ \mbox{and}\ \ \Col \sim (\RaFS)^{-1/5}\ \ \mbox{(rapid rotation)}.\label{equ:Colscaling}
\end{eqnarray}
Here the convective length scale is estimated from the power spectrum
of velocity, $E(k)$, for which $\uuu^2=\int E(k) dk$, either by taking
the wavenumber where the power has its maximum ($\kmax$) or the mean
wavenumber $\kmean = \int k E(k)dk/\int E(k)dk$. The length scales
$\lmax$ and $\lmean$ are given by $\ell_{\rm max,mean} =
\Lx/\tilde{k}_{\rm max,mean}$, where $\tilde{k}_{\rm max,mean} =
k_{\rm max,mean}/\kH$, and where $\kH=2\pi/\Lx=\pi/2d$ is the
wavenumber corresponding to the horizontal extent of the simulation
domain. \Figu{fig:kmaxmmean} shows $\tkmax$ and $\tkmean$ from
Set~A. For slow rotation ($\Co\lesssim 1$) both $\tkmax$ and $\tkmean$
are approximately constant, although the former is already consistent
with the $\Co^{1/2}$ scaling due to the large error estimates which
are taken to be the standard deviation of the mean values calculated
from several snapshots. For rapid rotation the $\Co^{1/2}$ scaling
from the CIA balance is recovered for both $\tkmax$ and $\tkmean$.

\begin{figure}[t]
  \centering
  \includegraphics[scale=.5]{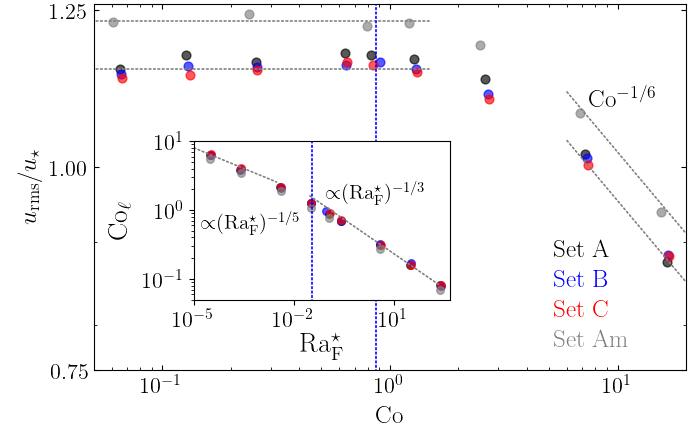}
  \caption{Volume-averaged rms-velocity as a function of $\Co$ from
    simulations in Sets~A, B, C, and Am from
    \cite{2023arXiv231012855K}. The dotted lines are either constant
    (for $\Co \le 1.5$) or proportional to $\Co^{-1/6}$ (for $\Co \ge
    6$). The inset shows $\Col$ as a function of $\RaFS$ for the same
    runs with power laws proportional to $(\RaFS)^{-1/5}$ for fast and
    $(\RaFS)^{-1/3}$ for slow rotation corresponding to
    $\RaFS\le3\cdot10^{-3}$ and $\RaFS\ge0.03$, respectively. The blue
    dotted vertical lines indicate the solar values of $\Co$ and
    $\RaFS$, respectively.}
  \label{fig:plot_urmscol_iau}
\end{figure}

The case that the deep parts of the solar convection zone is strongly
rotationally constrained has been discussed recently by \cite{FH16b}
and \cite{2021PNAS..11822518V}. Both of these studies argue that the
maximum horizontal scale of convection is reduced by rotation in the
deep parts of the convection zone, such that the largest convectively
driven scale coincides with supergranules (20--30~Mm) at a spherical
harmonic degree $\ell\approx100$. Using \Eq{equ:CoF} it is possible to
compute $\CoF$ at the base of the solar convection zone with $\Omsun
= 2.7\cdot 10^{-6}$~s$^{-1}$, $\Hp^\odot\approx 5\cdot 10^7$~m,
$\rho^\odot\approx 200$~kg~m$^{-3}$, and $\Ftot^\odot = \Lsun/(4\pi
\rbot^2)$, where $\Lsun = 3.83\cdot 10^{26}$~W, and $\rbot =
0.7\Rsun$, gives $\CoF^\odot \approx 3.1$. On the other hand,
\begin{eqnarray}
\Co = \frac{\ustar}{\urms}\frac{\CoF}{k_1\Hp}.
\end{eqnarray}
For the current slowly rotating simulations in Sets~A to C,
$\ustar/\urms \approx 0.87$ (see, \Fig{fig:plot_urmscol_iau}), and
$(k_1 \Hp)^{-1}\approx 0.32$, such that the solar $\CoF$ is achieved
in a simulation with $\Co\approx0.87$. Inspection of
\Fig{fig:kmaxmmean} suggests that the Sun is somewhere in between the
weakly rotationally influenced and the rotationally constrained
regimes.

The dominant convective scale in a simulation with $\Co\approx0.83$
and $\CoF=3.1$ is again estimated from the power spectrum of the
velocity. In this case the maximum power occurs at wavenumber $\tkmax
=3$ and the mean wavenumber is $\tkmean =7$, corresponding to length
scales $\lmax = 1.33d$ and $\lmean = 0.57d$. The pressure scale height
at the base of of the convective layer in this run is $\Hp =
0.49d$. Assuming the simulations to represent the deep parts of the
convection zone at the interface to the radiative layer, the pressure
scale height corresponds to $\Hp^\odot\approx 5\cdot 10^7$~m. This
leads to $\lmax \approx 135$~Mm and $\lmean \approx 58$~Mm,
respectively. These results seem to refute the idea that rotationally
constrained convection can explain the supergranular scale as the
largest convectively driven scale. Furthermore,
\cite{2023arXiv231012855K} showed that in the simulations of
\cite{FH16b}, where the supergranular scale is the dominant scale
correspond to a value of $\CoF$ that requires a rotation rate which is
at least 15 times higher than in the Sun.

\Fig{fig:plot_urmscol_iau} shows the time and volume-averaged
rms-velocity from Sets~A, B, C, and Am normalized by $\ustar$. The
scalings for slow and rapid rotation from \Eq{equ:uscaling} are
recovered for $\Co\lesssim 1.5$ and $\Co \gtrsim 6$, respectively. For
$\Co$ exceeding the maximum values here ($\Co\approx17$), the flow
begins to develop a large-scale vortical component \citep[see
  also][]{Chan07,2011ApJ...742...34K} which is likely due to
two-dimensionalization of turbulence, and extending the calculations
to higher $\Co$ becomes challenging. Finally, the local Coriolis
number $\Col$ is shown in the inset of
\Fig{fig:plot_urmscol_iau}. $\Col$ adheres to the scalings given in
\Eq{equ:Colscaling} with respect to $\RaFS$ for both slowly and
rapidly rotating regimes.

\section{Deardorff layer as a function of rotation}

In the canonical models of stellar convection relying on mixing length
theory \citep[e.g.][]{BV58}, the whole convection zone is unstably
stratified and convection is thought to be driven locally by a
superadiabatic temperature gradient
\begin{eqnarray}
\Delta\nabla = \nabla - \nabad = - \frac{1}{\Hp}\frac{{\rm d}s}{{\rm d}z} > 0,
\end{eqnarray}
where $\nabla = {\rm d}\ln \mean{T}/{\rm d} \ln p$ is the logarithmic
temperature gradient, $\nabad = ({\rm d}\ln \mean{T}/{\rm d} \ln
p)_{\rm ad} = 1 - 1/\gamma$ is the corresponding adiabatic gradient,
and where $\gamma = \cP/\cV$. If this were the case in the Sun, giant
cell convection on the scale of 200~Mm is expected to be
prominent. This is not observed in the Sun and 3D hydrodynamic
simulations suggest that the deep parts of convective layers are often
weakly subadiabatic
\cite[e.g.][]{1993A&A...277...93R,2015ApJ...799..142T,2017ApJ...845L..23K,2017ApJ...843...52H}. This
is thought to be due to the inherently non-local nature of convection
which is driven by cooling near the surface instead of heating from
the base as has been shown, e.g., in \cite{2017ApJ...845L..23K}.

\begin{figure}[t]
  \centering
  \includegraphics[scale=.38]{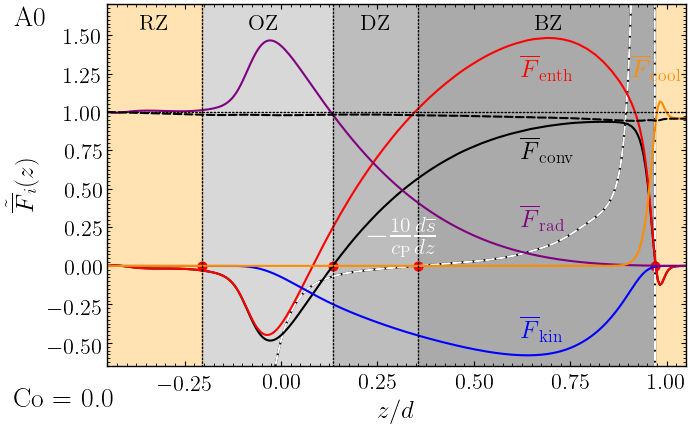}\includegraphics[scale=.38]{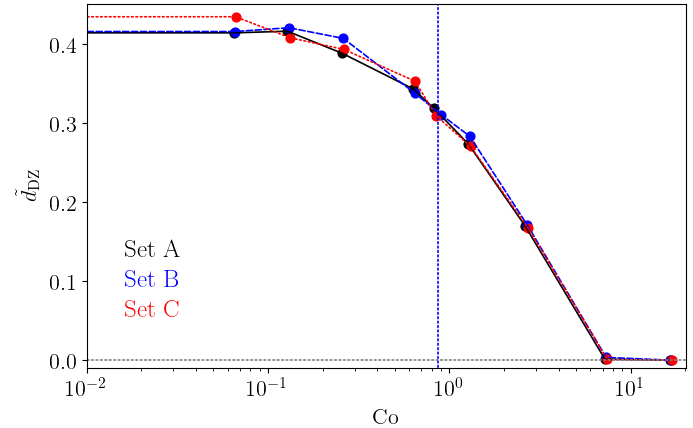}
  \caption{Left panel: Horizontally averaged energy fluxes from the
    non-rotating Run~A0 from \cite{2023arXiv231012855K}. Grey (orange)
    areas indicate mixed (radiative) layers, and the red circles
    highlight the boundaries between the layers. The various zones are
    characterised as $\mFrad\approx\Ftot$ (radiative; RZ),
    $\mFconv<0$, $\mFrad > \Ftot$ (overshoot; OZ), $\mFconv>0$,
    $\Delta\nabla < 0$ (Deardorff; DZ), and $\mFconv>0$, $\Delta\nabla
    > 0$ (buoyancy; BZ). Right: Depth of the Deardorff layer ($\ddz$)
    as a function of $\Co$ from simulations in Sets~A (black solid
    line), B (blue dashed), and C (red dotted). The blue dotted
    vertical line indicates the value of $\Co$ corresponding to the
    solar $\CoF$. Adapted from \cite{2023arXiv231012855K}.}
  \label{fig:plot_dear_iau}
\end{figure}

This subadiabatic but convecting layer is referred to as the Deardorff
zone (DZ), and it is characterised by $\Delta\nabla < 0$ and $\mFconv
> 0$, where
\begin{eqnarray}
\mFconv = \mFenth + \mFkin = \cP \mean{(\rho u_z)' T'} + \onehalf \mean{\rho \uuu^2 u_z},
\end{eqnarray}
is the total convected flux \citep[e.g.][]{CBTMH91}, which is the sum
of the time and horizontal averages of the enthalpy and kinetic energy
fluxes, and where the primes denote deviations from the horizontal
average. The left panel of \Fig{fig:plot_dear_iau} shows the flux
balance from a non-rotating Run~A0 from \cite{2023arXiv231012855K},
where in addition to $\mFconv$, $\mFenth$, and $\mFkin$, also the
radiative ($\mFrad = - \mean{K} {\rm d} \mean{T}/{\rm d} z $) and
cooling ($\mFcool = - \int \mean{\mathcal{C}} {\rm d}z$) fluxes, as
well as a quantity proportional to $\Delta \nabla$ are shown for
reference. The DZ in this run is over 40 per cent of the pressure
scale height at the base of the convection zone which is here defined
as the lower boundary of the DZ. For the run closest to the solar
value of $\CoF$, $\ddz \approx 0.3\Hp$ which corresponds to
$15$~Mm in the Sun.

\Fig{fig:plot_dear_iau} shows $\ddz$ as a fraction of the pressure
scale height at the base of the convection zone as a function of
rotation for Sets~A, B, and C from \cite{2023arXiv231012855K}. The
sets differ from each other in that the input energy flux is varied
such that between the extreme cases (Sets~A and C), $\Fn$ decreases by
factor of five. This has implications for the Mach number and also for
the overshooting below the convection zone
\citep[e.g.][]{2019A&A...631A.122K}. However, the depth of the DZ is
virtually unaffected by the change of $\Fn$. This is because the
cooling time at the surface is varied inversely proportional to $\Fn$
such that the thermal forcing remains unaffected.

\section{Effects of surface forcing}

There are some a few into the effects of varying surface forcing using
convection simulations. For example, \cite{CR16} varied the
superadiabatic gradient at the surface and found that it had a
substantial effect on the convective length scale and deep convection
zone dynamics. On the other hand, \cite{2019SciA....5.2307H} found
only a weak influence of the surface in a simulation that encompassed
nominally the entire convection zone of the Sun.

Here a similar approach as in \cite{CR16} is explored with a
simulation set-up that was used in \cite{2017ApJ...845L..23K}. In
distinction to the simulations discussed above, the upper cooling
layer is replaced by an imposed entropy gradient at the upper
boundary, and the $z$-coordinate runs between $-0.5\leq z/d \leq 1$
such that the transition between the initially isentropic and
radiative layers is at $z = 0$. Furthermore, the SGS diffusion term
has an extra term proportional to the mean entropy gradient:
\begin{eqnarray}
\FFF_{\rm SGS} = - (\chiSGS \rho T \bm\nabla s' + \chiSGS^{\rm m} \rho T \bm\nabla \mean{s}),\label{equ:FSGS2}
\end{eqnarray}
where $\chiSGS^{\rm m}$ is non-zero only above $z/d = 0.95$, such that
the second term on the rhs of \Eq{equ:FSGS2} transports the heat flux
through the upper boundary. Here $\PraSGS=1$ and $\PraSGS^{\rm m} =
\nu/\chiSGS^{\rm m} = 0.5$. In \cite{2017ApJ...845L..23K}, the entropy
gradient at the surface was fixed to $\widetilde{\nabla s} =
(d/\cP)(\eee_z \bm\cdot \bm\nabla \mean{s})=-10$. Here four values
between $-1$ and $-10$ for $\widetilde{\nabla s}$ are explored in
Set~G. In distinction to \cite{CR16} where a spatially fixed Newtonian
cooling term was used which does not allow the depth of the convection
zone to change appreciably, the current simulations use Kramers
opacity law which enables this.

\begin{figure}[t]
  \centering
  \includegraphics[scale=.38]{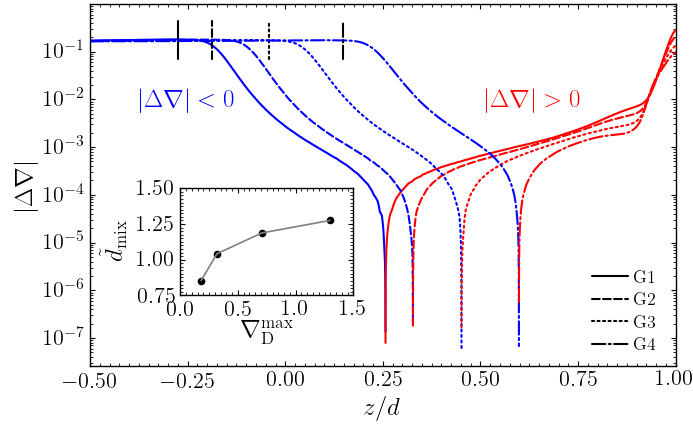}\includegraphics[scale=.38]{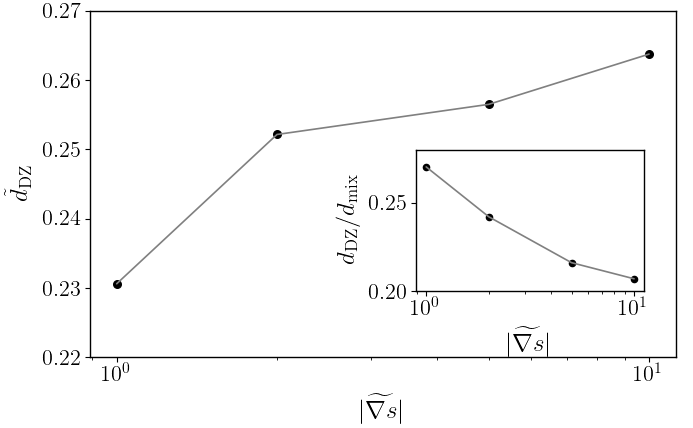}
  \caption{Left panel: Absolute value of the superadiabatic
    temperature gradient $\Delta\nabla$ for Runs~G[1-4]. Red (blue)
    parts of the curves indicate regions where $\Delta\nabla>0$
    ($\Delta\nabla<0$). The black vertical lines denote the bottom of
    the convectively mixed layer. The inset shows the depth of the
    mixed zone $\tdmix=\dmix/d$ as a function $\nabD^{\rm
      max}$. Right: depth of the Deardorff zone, $\tddz = \ddz/d$ as a
    function of $|\widetilde{\nabla s}|$. The inset shows the relative
    fraction of the Deardorff zone of the mixed zone as a function of
    $|\widetilde{\nabla s}|$.}
  \label{fig:plot_superad}
\end{figure}

In the absence of convection the hydrostatic solution with the Kramers
opacity law is convectively unstable only in a shallow surface layer
\citep[e.g.][]{BB14,2023arXiv231012855K}. This solution is modified by
the onset of convection and the final outcome is expected to be
sensitive to the surface physics. Decreasing $|\widetilde{\nabla s}|$
leads to a shallower convection zone as can be seen from the left
panel of \Fig{fig:plot_superad}. Furthermore, $\Delta\nabla$ near the
surface decreases and the surface temperature increases. In the
updated mixing length model of \cite{Br16}, the enthalpy flux was
quantified in terms of gradient ($\mFG$) and non-gradient ($\mFD$)
contributions
\begin{eqnarray}
\mFenth = \mFG +  \mFD = - \tau \mean{\rho} \mean{T} (\onethird \urms^2 \nabla_z \mean{s} + \mean{s'^2} g /\cP) = \onethird \mean{\rho} \cP \mean{T} (\tau \urms^2/\Hp) (\Delta\nabla + \nabD),
\end{eqnarray}
where $\tau$ is a relaxation time, and where the magnitude of the
non-gradient is characterised by
\begin{eqnarray}
\nabD = (3/\gamma)(\mean{s'^2}/\cP^2) \Ma^{-2}.
\end{eqnarray}
\cite{Br16} argued that $\nabD$ in deeper parts is proportional to its
maximum value near the surface, $\nabD^{\rm max}$. The depth of the
mixed layer $\dmix$, consisting of the buoyancy, Deardorff, and
overshoot zones, is shown as a function of $\nabD^{\rm max}$ in the
inset of the left panel of \Fig{fig:plot_superad} for the runs in
Set~G. While $\dmix$ increases monotonically with $\nabD^{\rm max}$
(corresponding to increasing $|\widetilde{\nabla s}|$), there appears
to be no straightforward relation between the two. The depth of the
Deardorff zone is not very sensitive to $|\widetilde{\nabla s}|$,
although its size relative to the mixed zone decreases somewhat as
$|\widetilde{\nabla s}|$ increases; see the inset of the right panel
of \Fig{fig:plot_superad}.

\section{Conclusions}

Several ways to address the ``convective conundrum,'' or the
discrepancy between convective velocity amplitudes in simulations and
solar observations, were reviewed based on results from recent
hydrodynamic Cartesian convection simulations. First the effects of
rotation from \cite{2023arXiv231012855K} were considered. These
results suggest that the convective scale in the deep convection zone
of the Sun is not sufficiently affected by rotation to reduce the
largest convectively driven scale to the supergranular scale of
$20$-$30$~Mm as has been conjectured earlier
\citep{FH16b,2021PNAS..11822518V}. These simulations also suggest that
the depth of the convective but formally stably stratified Deardorff
zone is reduced as rotation increases, but that a substantial
subadiabatic layer of about 15~Mm is still expected to be found at the
base of the solar convection zone. Scaling laws of several dynamical
quantities such as convective scale, velocity amplitude, and local
Coriolis number were shown to follow scalings derived under the CIA
balance
\citep[e.g.][]{1979GApFD..12..139S,2014ApJ...791...13B,2020PhRvR...2d3115A}.

The effects of surface forcing were explored with a set of new
simulations where the entropy gradient at the surface was imposed
similarly as in \cite{2017ApJ...845L..23K}. Unlike in the previous
studies in the literature that study the effects of the surface for
the deep convection zone, the current simulations allow the depth of
the convective layer to vary self-consistently. These preliminary
results show that stronger surface forcing, in terms of a steeper
entropy gradient, leads to a deeper convectively mixed layer. Although
there is a monotonic dependence between the imposed entropy gradient
and the depth of the convective layer, no clear relation between the
two can be identified. However, the fraction of the Deardorff layer of
the total depth of the convectively mixed layer decreases somewhat
when the surface forcing is increased.

The results quoted above come with the caveat that the surface forcing
of convection in the current simulations is assumed to be accurately
modelled. This, however, cannot be guaranteed, and it is likely that
much smaller scales need to be resolved to capture the effects of
radiative cooling in the photosphere accurately
\citep[e.g.][]{2017LRCA....3....1K}. Furthermore, the effects of
astrophysically relevant low Prandtl numbers
\citep[e.g.][]{1962JGR....67.3063S,2021A&A...655A..78K} and vigorous
magnetism \citep[e.g.][]{2022ApJ...933..199H} are also likely to play
important roles for solar and stellar convection.  \\
\\
{\bf Acknowledgements:} The simulations were performed using the
resources granted by the Gauss Center for Supercomputing for the
Large-Scale computing project ``Cracking the Convective Conundrum'' in
the Leibniz Supercomputing Centre's SuperMUC-NG supercomputer in
Garching, Germany. This work was supported in part by the Deutsche
Forschungsgemeinschaft (DFG, German Research Foundation) Heisenberg
programme (grant No.\ KA 4825/4-1) and by the Munich Institute for
Astro-, Particle and BioPhysics (MIAPbP) which is funded by the DFG
under Germany´s Excellence Strategy – EXC-2094 – 390783311.


\end{document}